\RequirePackage[mathlines]{lineno} % Display line numbers
\documentclass[a4paper,11pt]{article}
\usepackage{jheppub} % for details on the use of the package, please
\usepackage{overpic}
\usepackage{amssymb}
\usepackage{amsmath}
\usepackage[T1]{fontenc} % if needed
\usepackage{amsmath}
\usepackage{graphicx}
\usepackage{subfigure}
\usepackage{epstopdf}
\usepackage{rotating}
\usepackage{adjustbox} %DIF > 

\usepackage{threeparttable}%This package is used for tablenotes
\usepackage{multirow}
\usepackage{subfigure}
\usepackage{float}
\usepackage{verbatim}

\let\oldequation\equation
\let\oldendequation\endequation
\renewenvironment{equation}
  {\linenomathNonumbers\oldequation}
  {\oldendequation\endlinenomath}

\begin{document}
%\linenumbers

\title{\bf \boldmath
  Measurement of the branching fractions of doubly Cabibbo-suppressed $D$ decays
}

\collaborationImg{\includegraphics[height=4cm,angle=0 ]{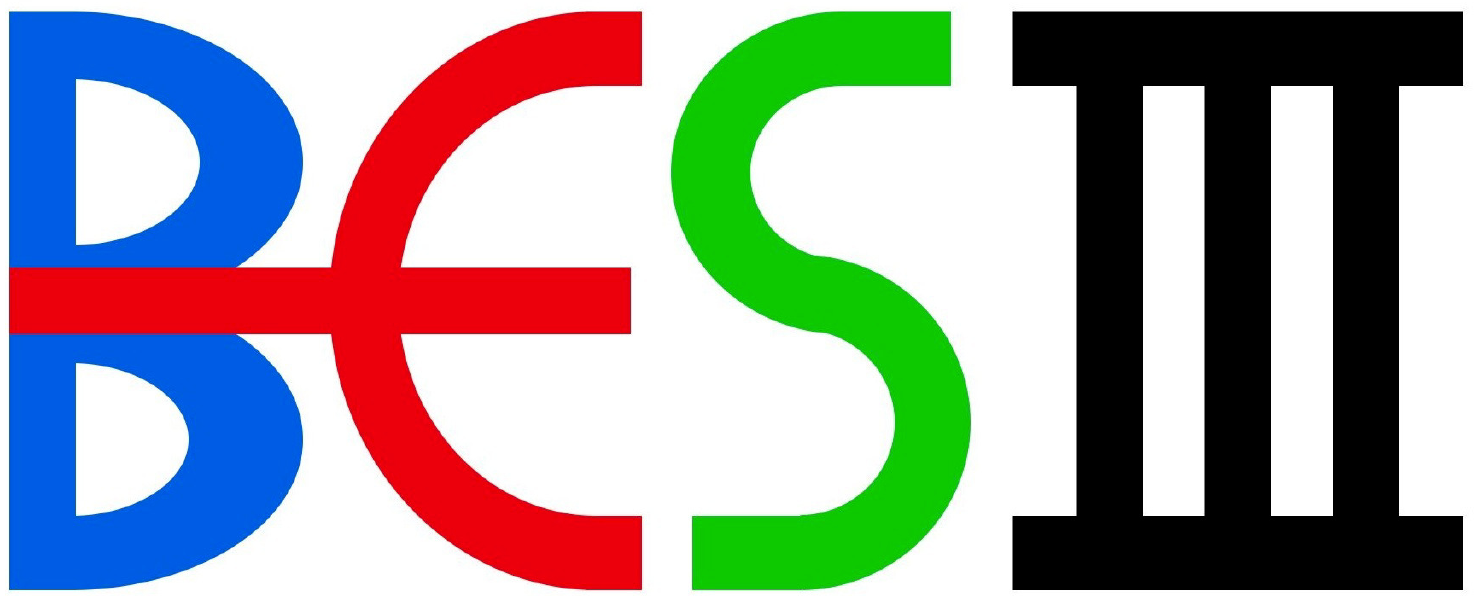}}
\collaboration{The BESIII Collaboration}
\emailAdd{besiii-publications@ihep.ac.cn}

%% Saved at => 2024-06-18
\author{
M.~Ablikim$^{1}$, M.~N.~Achasov$^{4,c}$, P.~Adlarson$^{76}$, O.~Afedulidis$^{3}$, X.~C.~Ai$^{81}$, R.~Aliberti$^{35}$, A.~Amoroso$^{75A,75C}$, Y.~Bai$^{57}$, O.~Bakina$^{36}$, I.~Balossino$^{29A}$
% ...
\\

{\it
~\\
$^{1}$ Institute of High Energy Physics, Beijing 100049, People's Republic of China\\
$^{2}$ Beihang University, Beijing 100191, People's Republic of China\\
$^{3}$ Bochum  Ruhr-University, D-44780 Bochum, Germany\\
$^{4}$ Budker Institute of Nuclear Physics SB RAS (BINP), Novosibirsk 630090, Russia\\
$^{5}$ Carnegie Mellon University, Pittsburgh, Pennsylvania 15213, USA\\
$^{6}$ Central China Normal University, Wuhan 430079, People's Republic of China\\
$^{7}$ Central South University, Changsha 410083, People's Republic of China\\
$^{8}$ China Center of Advanced Science and Technology, Beijing 100190, People's Republic of China\\
$^{9}$ China University of Geosciences, Wuhan 430074, People's Republic of China\\
$^{10}$ Chung-Ang University, Seoul, 06974, Republic of Korea\\
$^{11}$ COMSATS University Islamabad, Lahore Campus, Defence Road, Off Raiwind Road, 54000 Lahore, Pakistan\\
$^{12}$ Fudan University, Shanghai 200433, People's Republic of China\\
$^{13}$ GSI Helmholtzcentre for Heavy Ion Research GmbH, D-64291 Darmstadt, Germany\\
$^{14}$ Guangxi Normal University, Guilin 541004, People's Republic of China\\
$^{15}$ Guangxi University, Nanning 530004, People's Republic of China\\
$^{16}$ Hangzhou Normal University, Hangzhou 310036, People's Republic of China\\
$^{17}$ Hebei University, Baoding 071002, People's Republic of China\\
$^{18}$ Helmholtz Institute Mainz, Staudinger Weg 18, D-55099 Mainz, Germany\\
$^{19}$ Henan Normal University, Xinxiang 453007, People's Republic of China\\
$^{20}$ Henan University, Kaifeng 475004, People's Republic of China\\
$^{21}$ Henan University of Science and Technology, Luoyang 471003, People's Republic of China\\
$^{22}$ Henan University of Technology, Zhengzhou 450001, People's Republic of China\\
$^{23}$ Huangshan College, Huangshan  245000, People's Republic of China\\
$^{24}$ Hunan Normal University, Changsha 410081, People's Republic of China\\
$^{25}$ Hunan University, Changsha 410082, People's Republic of China\\
$^{26}$ Indian Institute of Technology Madras, Chennai 600036, India\\
$^{27}$ Indiana University, Bloomington, Indiana 47405, USA\\
$^{28}$ INFN Laboratori Nazionali di Frascati , (A)INFN Laboratori Nazionali di Frascati, I-00044, Frascati, Italy; (B)INFN Sezione di  Perugia, I-06100, Perugia, Italy; (C)University of Perugia, I-06100, Perugia, Italy\\
$^{29}$ INFN Sezione di Ferrara, (A)INFN Sezione di Ferrara, I-44122, Ferrara, Italy; (B)University of Ferrara,  I-44122, Ferrara, Italy\\
$^{30}$ Inner Mongolia University, Hohhot 010021, People's Republic of China\\
$^{31}$ Institute of Modern Physics, Lanzhou 730000, People's Republic of China\\
$^{32}$ Institute of Physics and Technology, Peace Avenue 54B, Ulaanbaatar 13330, Mongolia\\
$^{33}$ Instituto de Alta Investigaci\'on, Universidad de Tarapac\'a, Casilla 7D, Arica 1000000, Chile\\
$^{34}$ Jilin University, Changchun 130012, People's Republic of China\\
$^{35}$ Johannes Gutenberg University of Mainz, Johann-Joachim-Becher-Weg 45, D-55099 Mainz, Germany\\
$^{36}$ Joint Institute for Nuclear Research, 141980 Dubna, Moscow region, Russia\\
$^{37}$ Justus-Liebig-Universitaet Giessen, II. Physikalisches Institut, Heinrich-Buff-Ring 16, D-35392 Giessen, Germany\\
$^{38}$ Lanzhou University, Lanzhou 730000, People's Republic of China\\
$^{39}$ Liaoning Normal University, Dalian 116029, People's Republic of China\\
$^{40}$ Liaoning University, Shenyang 110036, People's Republic of China\\
$^{41}$ Nanjing Normal University, Nanjing 210023, People's Republic of China\\
$^{42}$ Nanjing University, Nanjing 210093, People's Republic of China\\
$^{43}$ Nankai University, Tianjin 300071, People's Republic of China\\
$^{44}$ National Centre for Nuclear Research, Warsaw 02-093, Poland\\
$^{45}$ North China Electric Power University, Beijing 102206, People's Republic of China\\
$^{46}$ Peking University, Beijing 100871, People's Republic of China\\
$^{47}$ Qufu Normal University, Qufu 273165, People's Republic of China\\
$^{48}$ Renmin University of China, Beijing 100872, People's Republic of China\\
$^{49}$ Shandong Normal University, Jinan 250014, People's Republic of China\\
$^{50}$ Shandong University, Jinan 250100, People's Republic of China\\
$^{51}$ Shanghai Jiao Tong University, Shanghai 200240,  People's Republic of China\\
$^{52}$ Shanxi Normal University, Linfen 041004, People's Republic of China\\
$^{53}$ Shanxi University, Taiyuan 030006, People's Republic of China\\
$^{54}$ Sichuan University, Chengdu 610064, People's Republic of China\\
$^{55}$ Soochow University, Suzhou 215006, People's Republic of China\\
$^{56}$ South China Normal University, Guangzhou 510006, People's Republic of China\\
$^{57}$ Southeast University, Nanjing 211100, People's Republic of China\\
$^{58}$ State Key Laboratory of Particle Detection and Electronics, Beijing 100049, Hefei 230026, People's Republic of China\\
$^{59}$ Sun Yat-Sen University, Guangzhou 510275, People's Republic of China\\
$^{60}$ Suranaree University of Technology, University Avenue 111, Nakhon Ratchasima 30000, Thailand\\
$^{61}$ Tsinghua University, Beijing 100084, People's Republic of China\\
$^{62}$ Turkish Accelerator Center Particle Factory Group, (A)Istinye University, 34010, Istanbul, Turkey; (B)Near East University, Nicosia, North Cyprus, 99138, Mersin 10, Turkey\\
$^{63}$ University of Bristol, (A)H H Wills Physics Laboratory; (B)Tyndall Avenue; (C)Bristol; (D)BS8 1TL\\
$^{64}$ University of Chinese Academy of Sciences, Beijing 100049, People's Republic of China\\
$^{65}$ University of Groningen, NL-9747 AA Groningen, The Netherlands\\
$^{66}$ University of Hawaii, Honolulu, Hawaii 96822, USA\\
$^{67}$ University of Jinan, Jinan 250022, People's Republic of China\\
$^{68}$ University of Manchester, Oxford Road, Manchester, M13 9PL, United Kingdom\\
$^{69}$ University of Muenster, Wilhelm-Klemm-Strasse 9, 48149 Muenster, Germany\\
$^{70}$ University of Oxford, Keble Road, Oxford OX13RH, United Kingdom\\
$^{71}$ University of Science and Technology Liaoning, Anshan 114051, People's Republic of China\\
$^{72}$ University of Science and Technology of China, Hefei 230026, People's Republic of China\\
$^{73}$ University of South China, Hengyang 421001, People's Republic of China\\
$^{74}$ University of the Punjab, Lahore-54590, Pakistan\\
$^{75}$ University of Turin and INFN, (A)University of Turin, I-10125, Turin, Italy; (B)University of Eastern Piedmont, I-15121, Alessandria, Italy; (C)INFN, I-10125, Turin, Italy\\
$^{76}$ Uppsala University, Box 516, SE-75120 Uppsala, Sweden\\
$^{77}$ Wuhan University, Wuhan 430072, People's Republic of China\\
$^{78}$ Yantai University, Yantai 264005, People's Republic of China\\
$^{79}$ Yunnan University, Kunming 650500, People's Republic of China\\
$^{80}$ Zhejiang University, Hangzhou 310027, People's Republic of China\\
$^{81}$ Zhengzhou University, Zhengzhou 450001, People's Republic of China\\
~\\
}
~\\
{
$^{a}$ Deceased\\
$^{b}$ Also at the Moscow Institute of Physics and Technology, Moscow 141700, Russia\\
$^{c}$ Also at the Novosibirsk State University, Novosibirsk, 630090, Russia\\
$^{d}$ Also at the NRC "Kurchatov Institute", PNPI, 188300, Gatchina, Russia\\
$^{e}$ Also at Goethe University Frankfurt, 60323 Frankfurt am Main, Germany\\
$^{f}$ Also at Key Laboratory for Particle Physics, Astrophysics and Cosmology, Ministry of Education; Shanghai Key Laboratory for Particle Physics and Cosmology; Institute of Nuclear and Particle Physics, Shanghai 200240, People's Republic of China\\
$^{g}$ Also at Key Laboratory of Nuclear Physics and Ion-beam Application (MOE) and Institute of Modern Physics, Fudan University, Shanghai 200443, People's Republic of China\\
$^{h}$ Also at State Key Laboratory of Nuclear Physics and Technology, Peking University, Beijing 100871, People's Republic of China\\
$^{i}$ Also at School of Physics and Electronics, Hunan University, Changsha 410082, China\\
$^{j}$ Also at Guangdong Provincial Key Laboratory of Nuclear Science, Institute of Quantum Matter, South China Normal University, Guangzhou 510006, China\\
$^{k}$ Also at MOE Frontiers Science Center for Rare Isotopes, Lanzhou University, Lanzhou 730000, People's Republic of China\\
$^{l}$ Also at Lanzhou Center for Theoretical Physics, Lanzhou University, Lanzhou 730000, People's Republic of China\\
$^{m}$ Also at the Department of Mathematical Sciences, IBA, Karachi 75270, Pakistan\\
$^{n}$ Also at Ecole Polytechnique Federale de Lausanne (EPFL), CH-1015 Lausanne, Switzerland\\
$^{o}$ Also at Helmholtz Institute Mainz, Staudinger Weg 18, D-55099 Mainz, Germany\\
}

%% ends here %%
}

\abstract{
  By analyzing $e^+e^-$ collision data collected at the center-of-mass energy of 3.773~GeV with the BESIII detector, corresponding to an integrated luminosity of 20.3~fb$^{-1}$, 
  we measure the branching fractions of the doubly Cabibbo-suppressed (DCS) decays
  $D^0\to K^+\pi^-$,
  $D^0\to K^+\pi^-\pi^-\pi^+$,
  $D^0\to K^+\pi^-\pi^0$,
  $D^0\to K^+\pi^-\pi^0\pi^0$,
  $D^+\to K^+\pi^+\pi^-$,
  and $D^+\to K^+K^+K^-$.  
  We also perform the first searches for
  $D^0\to K^+\pi^-\eta$,
  $D^0\to K^+\pi^-\pi^0\eta$,
  $D^+\to K^+\pi^+\pi^-\eta$,
  $D^{+} \to  K^{+} 
  \left(\pi^{+} \pi^{-} \eta\right)_{{\rm non}-\eta^{\prime}}$,
  and $D^+\to K^+\eta\eta$ and report the first observations and evidence 
for some of these final states.  
  Combining the measurements with the world averages of the corresponding Cabibbo-favored (CF) decays,
  the ratios of the DCS/CF branching fractions
  are obtained.
For the  
$D^{+} \to  K^{+} 
  \left(\pi^{+} \pi^{-} \eta\right)_{{\rm non}-\eta^{\prime}}$ decay, the ratio is significantly larger than the corresponding ratios of the other DCS decays.
}
% \pacs{13.20.Fc, 14.40.Lb}

\maketitle
\flushbottom

\section{Introduction}

The study of the doubly Cabibbo-suppressed~(DCS) decays of charmed mesons is 
important for the  understanding of the decay dynamics of charm quarks.
It is expected that the branching fraction (BF) of the DCS $D$ decays is 
about $(0.5-2.0)$ ${\rm\tan}^4\theta_C$ times its Cabibbo-favored~(CF)
counterpart~\cite{Lipkin:2002za,Cheng:2010ry}, where ${\rm\theta}_{C}$ is the Cabibbo mixing angle~\cite{Cabibbo:1963yz}.
All known DCS $D$ decays approximately support this expectation~\cite{ParticleDataGroup:2024cfk}, except for the  recently observed $D^+\to K^+\pi^+\pi^-\pi^0$ decay.
This decay was observed for the first time by BESIII~\cite{BESIII:2020wnc,BESIII:2021uix} and later confirmed by Belle~\cite{Belle:2022aha};
the world average BF is $(1.10 \pm 0.07)\times 10^{-3}$, giving a DCS/CF BF ratio of $(6.11\pm0.42)\tan^4\theta_{C}$, which is significantly greater than the expectation.
Later, BESIII reported
the first observations of $D^+\to K^+\pi^0\pi^0$ and $D^+\to K^+\pi^0\eta$ decays~\cite{BESIII:2021qsa},
the measurement of $D^0\to K^+\pi^-\pi^0(\pi^0)$~\cite{BESIII:2022chs},
and the study of $D^+_s\to K^+K^+\pi^-(\pi^0)$~\cite{BESIII:2023coc}.
Charge conjugate decays are implied throughout this paper.

In this paper, we present the measurement of the BFs of the DCS decays 
$D^0\to K^+\pi^-$,
$D^0\to K^+\pi^-\pi^-\pi^+$,
$D^0\to K^+\pi^-\pi^0$,
$D^0\to K^+\pi^-\pi^0\pi^0$,
$D^0\to K^+\pi^-\eta$,
and $D^0\to K^+\pi^-\pi^0\eta$ by using the semileptonic $\bar D^0\to K^+ e^-\bar \nu_e$ and $\bar D^0\to K^+ \mu^-\bar \nu_\mu$ tags, and
$D^+\to K^+\pi^+\pi^-$,
$D^+\to K^+K^+K^-$,
$D^+\to K^+\pi^+\pi^-\eta$,
$D^{+} \to  K^{+} %\pi^{+} \pi^{-} \eta_{{\rm non}-\eta'}
  \left(\pi^{+} \pi^{-} \eta\right)_{{\rm non}-\eta^{\prime}}$,
and $D^+\to K^+\eta\eta$ by using the hadronic $D^- \to K^{+}\pi^{-}\pi^{-}$,
$D^- \to K^0_{S}\pi^{-}$ and $D^-\to K^{+}\pi^{-}\pi^{-}\pi^{0}$ tags.
Combining the obtained values with the world averages of their corresponding CF decays,
the ratios of the DCS/CF branching fractions $\mathcal B_{\rm DCS}/\mathcal B_{\rm CF}$ are obtained.
For the $D^0$ decays, the CF and DCS decays into the same final state are indistinguishable, causing a much higher background when using hadronic tags. 
In addition, the use of hadronic tags suffers from complicated corrections due to quantum correlation,
especially for those decays with no knowledge of the strong-phase differences~\cite{ParticleDataGroup:2024cfk}.
For these reasons, a semileptonic tag method is adopted in this work to investigate the DCS $D^0$ decays.
We note that $D^0\bar D^0$ mixing effects are negligible with our precision. 

The measurements are performed using $e^+e^-$ collision data corresponding to an integrated luminosity of 20.3\,fb$^{-1}$~\cite{BESIII:2015equ,BESIII:2024lbn} collected at the center-of-mass energy of $\sqrt s=3.773$~GeV with the BESIII detector.

\section{Data and Monte Carlo Generation}

The BESIII detector~\cite{BESIII:2009fln} records symmetric $e^+e^-$ collisions
provided by the BEPCII storage ring~\cite{Yu:2016cof}
in the center-of-mass energy range from 1.85 to 4.95~GeV,
with a peak luminosity of $1.1 \times 10^{33}\;\text{cm}^{-2}\text{s}^{-1}$
achieved at $\sqrt{s} = 3.773\;\text{GeV}$.
BESIII has collected large data samples in this energy region~\cite{BESIII:2020nme,lu2020online,zhang2022suppression}. The cylindrical core of the BESIII detector covers 93\% of the full solid angle and consists of a helium-based
multilayer drift chamber~(MDC), a plastic scintillator time-of-flight
system~(TOF), and a CsI(Tl) electromagnetic calorimeter~(EMC),
which are all enclosed in a superconducting solenoidal magnet
providing a 1.0~T magnetic field.
The solenoid is supported by an
octagonal flux-return yoke with resistive plate counter muon
identification modules~(MUC) interleaved with steel.
The charged-particle momentum resolution at $1~{\rm GeV}/c$ is
$0.5\%$, and the
${\rm d}E/{\rm d}x$
resolution is $6\%$ for electrons
from Bhabha scattering. The EMC measures photon energies with a
resolution of $2.5\%$ ($5\%$) at $1$~GeV in the barrel (end cap)
region.
The time resolution in the TOF barrel region is 68~ps, while
that in the end cap region was 110~ps.
The end cap TOF
system was upgraded in 2015 using multigap resistive plate chamber
technology, providing a time resolution of
60~ps,
which benefits 85\% of the data used in this analysis
~\cite{li2017study,guo2017study,CAO2020163053}.

Simulated samples produced with the {\sc geant4}-based~\cite{GEANT4:2002zbu} Monte Carlo (MC) package, which
includes the geometric description of the BESIII detector and the
detector response, are used to determine the detection efficiency
and to estimate the backgrounds. The simulation includes the beam-energy spread and initial-state radiation in the $e^+e^-$
annihilations modeled with the generator {\sc kkmc}~\cite{Jadach:2000ir,Jadach:1999vf}.
The inclusive MC samples consist of the production of $D\bar{D}$ pairs,
the non-$D\bar{D}$ decays of the $\psi(3770)$, the initial-state radiation
production of the $J/\psi$ and $\psi(3686)$ states, and the
continuum processes.
The known decay modes are modelled with {\sc
    evtgen}~\cite{Ping:2008zz,Lange:2001uf} using the BFs taken from the
Particle Data Group (PDG)~\cite{ParticleDataGroup:2024cfk}, and the remaining unknown decays of the charmonium states are
modeled by {\sc lundcharm}~\cite{Chen:2000tv}. Final-state radiation is incorporated using the {\sc photos} package~\cite{Richter-Was:1992hxq}.

To ensure consistency between the MC simulations and the data, and to better estimate the efficiencies, the well-known decays are generated combining different processes to decrease systematic uncertainties. The $D^0\to K^+\pi^-\pi^0$ decay is simulated using a generator which combines the resonant decays $D^0\to K^{*0}(892)\pi^0$, $D^0\to K^{*+}(892)\pi^-$, $D^0\to K^+\rho^-(770)$, and a phase-space model.
The $D^+\to K^+\pi^+\pi^-$ decay is simulated combining the resonant decays
$D^+\to K^{+}\rho^0(770)$, $D^+\to K^{*0}(892)\pi^+$, $D^+\to K^{+}f_{0}(980)$, and $D^+\to K_2^{*0}(1430)\pi^+$~\cite{ParticleDataGroup:2024cfk}.
The other DCS $D$ decays are simulated with a phase-space model.
The $\bar D^0 \to K^+\ell^- \bar\nu_\ell$ decay is simulated with
a two-parameter series expansion of the form factor~\cite{Becirevic:1999kt,BESIII:2015tql,Tchikilev:1999vpw}.

\section{Event selection}
\label{eventselection}

Charged tracks detected in the MDC (except for those used for $K^0_S$ reconstruction) are required to originate from a region within $|\rm{cos\theta}|<0.93$, $|V_{xy}|<$ 1\,cm and $|V_{z}|<$ 10\,cm.
Here, $\theta$ is the polar angle of the charged track with respect to the MDC axis, $|V_{xy}|$ and $|V_{z}|$ are the distances of closest approach of the charged track to the interaction point perpendicular to and along the MDC axis, respectively.
Particle identification~(PID) for charged tracks combines measurements of the energy deposited in the MDC~(d$E$/d$x$) and the flight time in the TOF to form likelihoods $\mathcal{L}(h)$, for each hadron hypothesis $h = K,\pi$. 
Charged tracks with  $\mathcal{L}(K)>\mathcal{L}(\pi)$ and $\mathcal{L}(\pi)>\mathcal{L}(K)$ are assigned as charged kaons and pions, respectively.

Each $K_{S}^0$ candidate is reconstructed from two oppositely-charged tracks satisfying $|V_{z}|<$ 20~cm.
The two charged tracks are assigned
as $\pi^+\pi^-$ without imposing PID criteria. They are constrained to
originate from a common vertex and are required to have an invariant mass
within $M_{\pi^{+}\pi^{-}} \in$ (0.487,0.511)~GeV$/c^{2}$.
The
decay length of the $K^0_S$ candidate is required to be greater than
twice the vertex resolution away from the interaction point.
Both the primary vertex fit and second vertex fit are required 
to have a $\chi^2 < 100$, and the updated $K^0_S$ four-vector 
is used for later kinematics.  

Photon candidates are selected from showers reconstructed in the EMC.
The shower time is required to be within 700\,ns of the event start time.
The shower energy is required to be greater than $25~\mathrm{MeV}$ in the barrel $(|\rm{cos\theta}|<0.80)$ and $50~\mathrm{MeV}$ in the end cap $(0.86<|\rm{cos\theta}|<0.92)$ regions.
The opening angle between the shower direction and the extrapolated position at the EMC of the closest  charged track must be greater than $10^{\circ}$.
The $\pi^0$ and $\eta$ candidates are formed by photon pairs with an invariant mass within $(0.115,\,0.150)$ and $(0.505,\,0.575)$\,GeV$/c^{2}$, respectively.To improve the resolution, a kinematic fit constraining the $\gamma\gamma$ invariant mass to the nominal $\pi^{0}$ or $\eta$ mass~\cite{ParticleDataGroup:2024cfk} is imposed on the selected photon pair.
The kinematic fit is required to have a $\chi^2 < 50$.
The four-momentum of the $\pi^0$ or $\eta$ candidate updated by this kinematic fit is kept for further analysis.

\section{Analysis of the DCS $\mathbf{D^0}$ decays}

At the center-of-mass energy of $\sqrt s=3.773$~GeV, the $D^0\bar D^0$ pairs are produced copiously and are not accompanied by additional hadrons.
This allows the use of the double-tag~(DT) method~\cite{MARK-III:1985hbd,MARK-III:1987jsm} to study the $D$ decays.
In this section, DT events refer to events in which the DCS $D^0$ decays are found recoiling against the semileptonic decays $\bar D^0\to K^+ e^-\bar \nu_e$ and $\bar D^0\to K^+ \mu^-\bar \nu_\mu$.
The BF of each DCS $D^0$ decay is determined by
\begin{equation}\label{equ:br}
  {\mathcal B}_{{\rm sig}} = \frac{N_{\rm DT}} {2\, N_{D^0\bar D^0}\, 
    \epsilon_{\rm DT}\,  {\mathcal B}_{\rm SL}},
\end{equation}
where $N_{\rm DT}$ is the number of the DT events observed in the data sample, 
$N_{D^0\bar D^0}=(73.29\pm0.84) \times 10^6$
is the total number of $D^0\bar D^0$ pairs in the data sample (calculated
by combining the $e^+e^-\to D^0\bar D^0$ cross section~\cite{BESIII:2018iev} and the integrated luminosity of the data sample~\cite{BESIII:2024lbn} at $\sqrt s=3.773$ GeV),
$\epsilon_{\rm DT}$ is the efficiency of reconstructing the DT events,
and
${\mathcal B}_{\rm SL}$ is the BF of $\bar D^0\to K^+ e^-\bar \nu_e$ or $\bar D^0\to K^+ \mu^-\bar \nu_\mu$ quoted from the PDG~\cite{ParticleDataGroup:2024cfk}.

The signal candidates for the DCS $D^0$ decays are identified by means of two variables: the energy difference
$\Delta E_{\rm sig} \equiv E_{D^0} - E_{\rm beam}$
and the beam-constrained mass
$M^{\rm sig}_{\rm BC} \equiv \sqrt{E^{2}_{\rm beam}-|\vec{p}_{D^0}|^{2}}$,
where $E_{\rm beam}$ is the beam energy, $\vec{p}_{D^0}$ and $E_{D^0}$ are the momentum and energy of the $D^0$ candidate in the $e^+e^-$ rest frame, respectively.
If there are multiple candidates for the hadronic side,
only the one with the minimum $|\Delta E_{\rm sig}|$ is kept.
Correctly reconstructed $D^0$ candidates concentrate around zero in the $\Delta E_{\rm sig}$ distribution
and around the nominal $D^0$ mass in the $M^{\rm sig}_{\rm BC}$ distribution.
Events satisfying the $\Delta E_{\rm sig}$ requirements for each signal decay, as
listed in Table~\ref{tab:D0_sigyield}, are kept for further analysis.

In the selection of the $D^0\to K^+\pi^-\pi^-\pi^+$ candidates,
if either combination of $\pi^+ \pi^-$ falls into the $K_S^0$ mass window $|M_{\rm\pi^+\pi^-}-0.50|<0.03$~GeV/$c^2$, the event is vetoed to reject the dominant peaking background from the singly Cabibbo-suppressed decay $D^0\to K^+\pi^-K_S^0(\to\pi^+\pi^-)$.
For $D^0\to K^+\pi^-\pi^0\pi^0$, we also veto candidates with $\pi^0\pi^0$ masses falling into $|M_{\rm\pi^0\pi^0}-0.50|<0.03$~GeV/$c^2$, to reject background from the $D^0\to K^+ \pi^- K_{S}^{0}(\to\pi^0\pi^0)$ decay.
Both $K_{S}^{0}$ veto requirements correspond to at least 3.5 standard deviations of the experimental $K_{S}^{0}$ mass resolution.

Once the hadronic decays of the $D^0$ mesons are reconstructed, the candidates for the $\bar D^0\to K^+e^-\bar \nu_{e}$ or the $\bar D^0\to K^+\mu^-\bar \nu_{\mu}$ decay  are selected from the remaining tracks that have not been used for the selection of the hadronic side.
The charged kaons from the semileptonic $\bar D^0$ decays are required to satisfy the same PID criteria as the kaons from the hadronic $D^0$ decays, and
the charge of the lepton candidate is required to be opposite to that of the kaon from the semileptonic $\bar D^0$ decay.
The number of additional charged tracks, $N_{\rm extra}^{\rm charge}$, is required to be zero.  
Electron PID uses the combined d$E$/d$x$, TOF, and EMC information to calculate confidence levels under the electron, pion, and kaon hypotheses, $\mathit{CL}_e$, $\mathit{CL}_{\pi}$, and $\mathit{CL}_{K}$.
Electron candidates are required to satisfy $\mathit{CL}_e>0.001$ and $\mathit{CL}_e/(\mathit{CL}_e+\mathit{CL}_\pi+\mathit{CL}_K)>0.8$.
To further reduce the background due to mis-identification between hadrons and electrons, the energy of the electron candidate deposited in the EMC is required to be greater than 0.8 times its momentum measured by the MDC. 
The four-momenta of the photon(s) within $5^\circ$ of the initial electron direction are added to the electron four-momentum in order to partially compensate the effects of the final-state radiation and bremsstrahlung (FSR recovery)~\cite{BESIII:2015tql}.  
Muon PID uses measurements from the d$E$/d$x$, TOF and EMC; the MUC is not used since most muons have momenta less than 0.5 GeV/$c$.  
Based on these measurements, we calculate the combined confidence levels for electron, muon, pion, and kaon hypotheses, $\mathit{CL}_e$, $\mathit{CL}_\mu$, $\mathit{CL}_\pi$, and $\mathit{CL}_K$.
The charged track satisfying $\mathit{CL}_\mu>0.001$, $\mathit{CL}_\mu > Max(\mathit{CL}_e, \mathit{CL}_\pi, \mathit{CL}_K)$, and $E_{\mathrm{EMC}} \in$ $(0.1,0.3)~\mathrm{GeV}$ is assigned as muon candidate, where $E_{\mathrm{EMC}}$ is its energy deposited in the EMC.

To suppress potential backgrounds from hadronic decays with a misidentified lepton,
the invariant mass of the $K^+\ell^-$ combination, $M_{K^+\ell^-}$, is required to be less than 1.8~GeV/$c^2$ for $\bar D^0\to K^+e^-\bar \nu_e$, and less than 1.5~GeV/$c^2$ for $\bar D^0\to K^+\mu^-\bar \nu_\mu$ .
Furthermore, we require that
the maximum energy of the additional photons ($E^{\rm max}_{\rm extra\,\gamma}$) which are not used in the tag
selection is less than 0.25~GeV and there is no additional good $\pi^0$ candidate ($N_{\rm extra}^{\pi^0}$).

\begin{figure*}[htbp]
  \centering
  \includegraphics[width=1.0\textwidth]{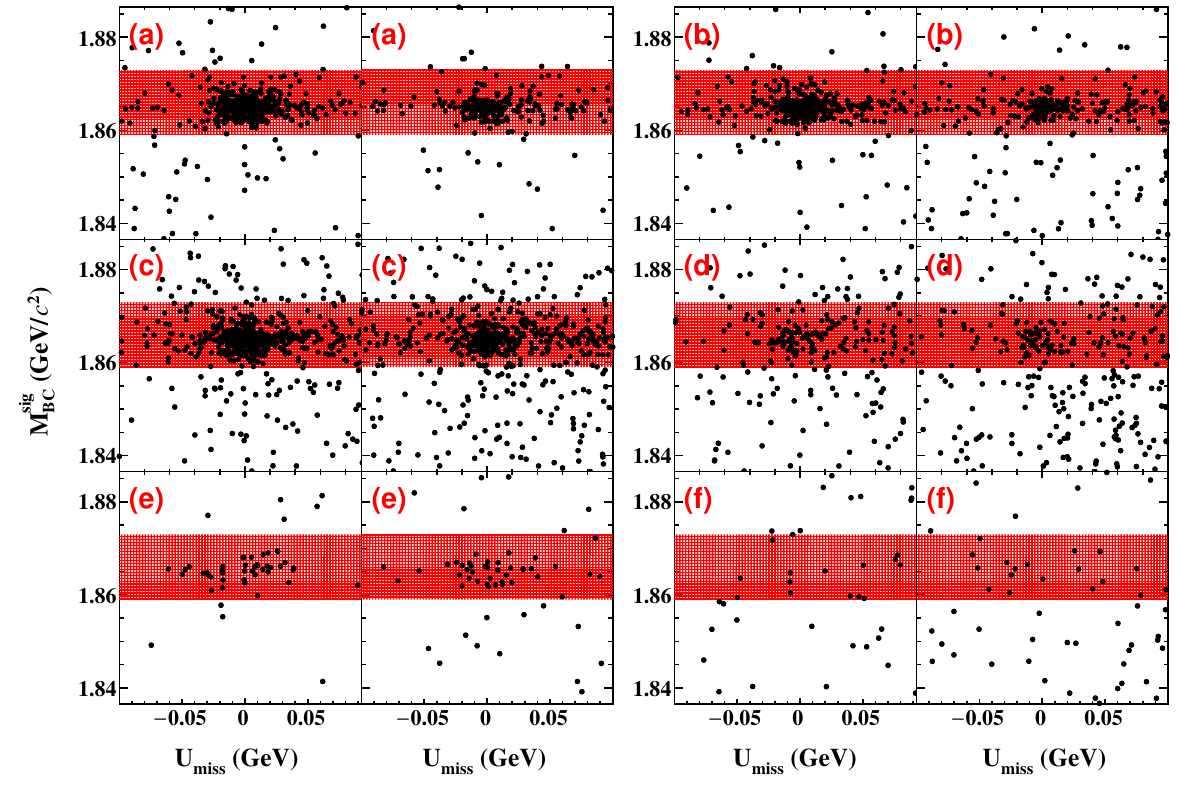}
  \caption{Distributions of $M^{\rm sig}_{\rm BC}$ versus $U_{\mathrm{miss}}$ of the DT candidate events for (a) $D^0\to K^+\pi^-$, (b) $D^0\to K^+\pi^-\pi^-\pi^+$, (c) $D^0\to K^+\pi^-\pi^0$, (d) $D^0\to K^+\pi^-\pi^0\pi^0$, (e) $D^0\to K^+\pi^-\eta$ and (f) $D^0\to K^+\pi^-\pi^0\eta$ in data.  
In each pair, the left plot is for $\bar D^0\to K^+ e^-\bar \nu_e$ tags and the right is for $\bar D^0\to K^+ \mu^-\bar \nu_\mu$ tags. 
    The red shaded areas show the $M^{\rm sig}_{\rm BC}$ signal regions.
  }
  \label{fig:M_BC_M2}
\end{figure*}

\begin{figure*}[htbp]
  \centering
  \includegraphics[width=1.0\textwidth]{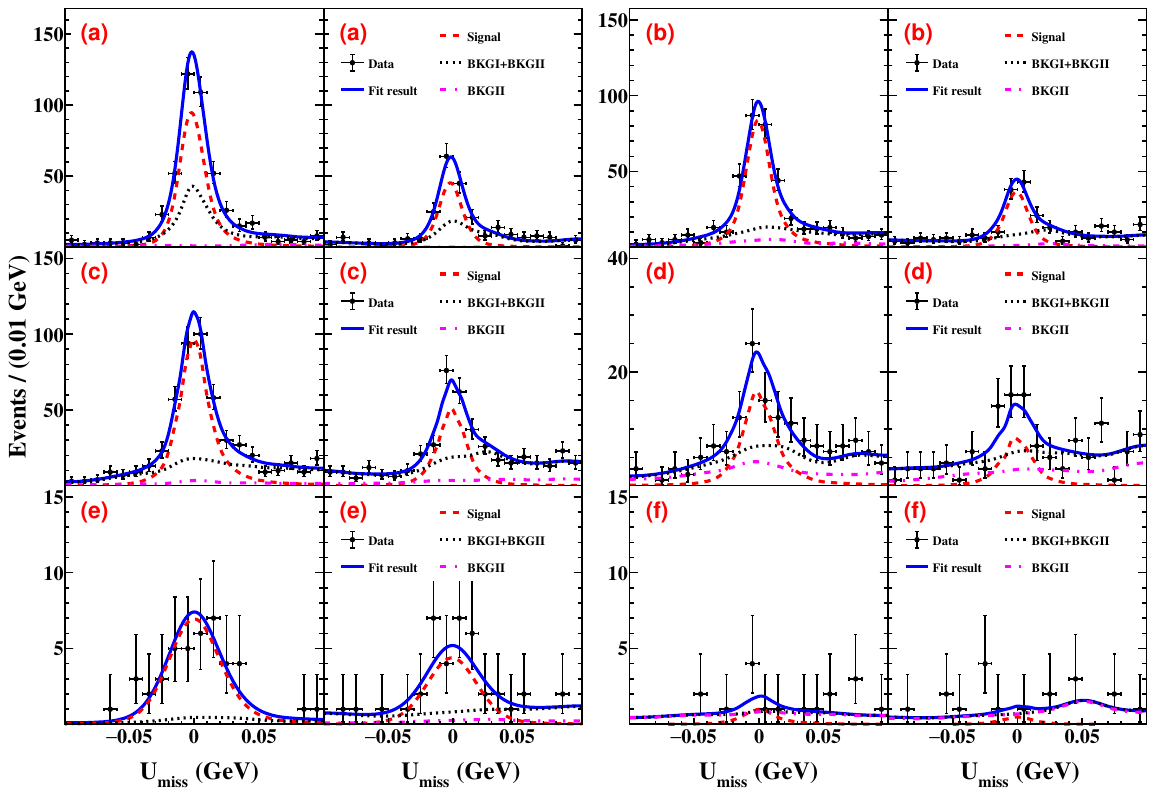}
  \caption{Fits to the $U_{\mathrm{miss}}$ distributions of the DT candidate events for (a) $D^0\to K^+\pi^-$, (b) $D^0\to K^+\pi^-\pi^-\pi^+$, (c) $D^0\to K^+\pi^-\pi^0$, (d) $D^0\to K^+\pi^-\pi^0\pi^0$, (e) $D^0\to K^+\pi^-\eta$ and (f) $D^0\to K^+\pi^-\pi^0\eta$ in data.  In each plot pair, the left panel is $\bar D^0\to K^+ e^-\bar \nu_e$ tags and the right one is $\bar D^0\to K^+ \mu^-\bar \nu_\mu$ tags.
    The points with error bars are data and the solid blue curves are the total fit results.  
The dotted red curves are the fitted signal portion, the dashed black curves are the summed BKGI+BKGII background components, and the dot-dashed pink curves are the BKGII contributions.
  }
  \label{fig:fits_umiss}
\end{figure*}

The semileptonic $\bar D^0$ decay is identified using the variable
$ U_{\mathrm{miss}}\equiv E_{\mathrm{miss}}-|\vec{p}_{\mathrm{miss}}|$,
where $E_{\mathrm{miss}}\equiv E_{\mathrm{beam}}-E_{K^+}-E_{\ell^-}$ and $\vec{p}_{\mathrm{miss}}\equiv
  \vec{p}_{\bar D^0}-\vec{p}_{K^+}-\vec{p}_{\ell^-}$ are the missing energy and momentum of the DT event in the $e^+e^-$ center-of-mass system, in
which $E_{K^+}$ and $\vec{p}_{K^+}$ are the energy and momentum of the $K^+$, and
$E_{\ell^-}$ and $\vec{p}_{\ell^-}$ are the energy and momentum of the $\ell^-$. The
$U_{\mathrm{miss}}$ resolution is improved by setting the $D^0$ energy to the beam energy and imposing $\vec{p}_{\bar D^0} \equiv {-\hat{p}_{D^0}} \, \sqrt{E_{\mathrm{beam}}^{2}-M_{D^0}^{2}}$, where $\hat{p}_{D^0}$ is the unit vector in the momentum direction of the $D^0$ and $M_{D^0}$ is the $D^0$ nominal mass~\cite{ParticleDataGroup:2024cfk}.

Figure \ref{fig:M_BC_M2} shows the distributions of $M^{\rm sig}_{\rm BC}$ versus $U_{\rm miss}$ of the accepted DT candidate events in data. The signal DT candidate events have $M^{\rm sig}_{\rm BC}$ values near the $D^0$ mass and $U_{\rm miss}$ values near zero.
The signal region in $M^{\rm sig}_{\rm BC}$ is chosen as $M^{\rm sig}_{\rm BC}\in (1.859,1.873)$\,GeV/$c^2$.
With the above-mentioned requirements, the $U_{\rm miss}$ distributions of the accepted events are shown in Fig.~\ref{fig:fits_umiss}.
The detection efficiencies $\epsilon_{{\rm DT}}$, estimated using the signal MC samples, are listed in Table~\ref{tab:D0_sigyield}.

The background components have been estimated by studying the inclusive MC sample, and are listed in Table~\ref{BKG1}. The largest contribution comes from CF decays due to particle misidentification~(BKGI),
with some remaining background contributions from other decays~(BKGII).
To extract the signal yields, simultaneous unbinned maximum-likelihood fits are performed to the $U_{\mathrm{miss}}$ distributions of the accepted DT candidate events tagged by $\bar D^0\to K^+ e^-\bar \nu_e$ and $\bar D^0\to K^+ \mu^-\bar \nu_\mu$.
For each signal decay, the simultaneous fit constrains the two semileptonic tags to share a common signal decay BF value.
The signal shapes are derived from the signal MC samples  convolved with a Gaussian function, and the background shapes from the inclusive MC sample.
The yield of the BKGI component is fixed based on the known BFs and the misidentification rates, while the yields of the signal and BKGII components are left free.
The fit results are shown in Fig.~\ref{fig:fits_umiss};
from the fit we obtain the BF values of each signal channel, listed in Table~\ref{tab:D0_sigyield}.
The statistical significance is estimated by evaluating the change in negative log-likelihood values when the signal is included or excluded in the fits, and calculating the related probability under the $\chi^2$ distribution hypothesis, accounting for the change in the number of degrees of freedom~\cite{Cowan:2010js}.

\begin{table*}[htpb]
  \centering
  \caption{The common $\Delta E_{\rm sig}$ requirements, separate detection efficiencies ($\epsilon^{e}_{\rm DT}$ for the electron mode and $\epsilon^{\mu}_{\rm DT}$ for the muonic mode) of the DCS $D^0$ decays, the separate ($N^{e}_{\rm DT}$ for the electron mode and $N^{\mu}_{\rm DT}$ for the muonic mode) and combined signal ($N_{\rm DT}$) yields in data, the obtained BFs, and the statistical significances for each signal decay. 
Here, separate and combined refer to the $\bar D^0\to K^+ e^-\bar \nu_e$ and $\bar D^0\to K^+ \mu^-\bar \nu_\mu$ tags, the efficiencies do not include the BFs of $\pi^0\to \gamma\gamma$ and $\eta\to \gamma\gamma$, and the uncertainties are statistical only.}
  \renewcommand\arraystretch{1.3}
  \adjustbox{width=\textwidth}{%
    \begin{tabular}{l|c|c|c|c|c|c|c|c}
      \hline\hline
      Signal decay                              & $\Delta E_{\rm sig}$ (MeV) & $\epsilon^{e}_{\rm DT}~(\%)$ & $\epsilon^{\mu}_{\rm DT}~(\%)$ & $N^{e}_{\rm DT}$       & $N^{\mu}_{\rm DT}$   & $N_{\rm DT}$            & $\mathcal B_{\rm sig}~(\times10^{-4})$ & Significance ($\sigma$) \\ \hline
$D^0 \to  K^{+} \pi^{-}$&$(-31,30)$&$38.53\pm0.15$&$18.79\pm0.12$&$259.2\pm17.7$&$121.7\pm8.3$&$381.0\pm26.0$&$1.30\pm0.09$&$>12.0$\\
$D^0 \to  K^{+} \pi^{-} \pi^{-}  \pi^{+}$&$(-31,29)$&$19.36\pm0.12$&$8.32\pm0.09$&$238.9\pm18.7$&$98.9\pm7.8$&$337.8\pm26.5$&$2.38\pm0.19$&$>12.0$\\
$D^0 \to  K^{+} \pi^{-} \pi^{0}$&$(-55,41)$&$19.92\pm0.13$&$9.71\pm0.09$&$312.6\pm21.2$&$146.7\pm9.9$&$459.3\pm31.1$&$3.06\pm0.21$&$>12.0$\\
$D^0 \to  K^{+} \pi^{-} \pi^{0}  \pi^{0}$&$(-62,42)$&$8.20\pm0.09$&$3.93\pm0.06$&$58.1\pm11.4$&$26.8\pm5.2$&$84.9\pm16.6$&$1.40\pm0.27$&5.0\\
$D^0 \to  K^{+} \pi^{-} \eta$&$(-35,36)$&$19.91\pm0.13$&$9.50\pm0.09$&$42.5\pm6.7$&$19.5\pm3.1$&$62.1\pm9.7$&$1.04\pm0.16$&5.8\\
$D^0 \to  K^{+} \pi^{-} \pi^{0}  \eta$&$(-47,38)$&$7.97\pm0.09$&$3.87\pm0.06$&$3.5\pm3.3$&$1.7\pm1.5$&$5.2\pm4.8$&$0.22\pm0.20$&0.3\\
      \hline\hline
    \end{tabular}%
  }
  \label{tab:D0_sigyield}
\end{table*}

\begin{table}[htpb]\centering
  \caption{The BKGI components for each signal decay from the CF decay due to particle misidentification.}
  \begin{tabular}{l|l}
    \hline\hline
  Signal                          & BKGI                            \\ \hline
  $D^0 \to K^+ \pi^-$             & $D^0 \to K^- \pi^+$             \\ \hline
  $D^0 \to K^+ \pi^- \pi^- \pi^+$ & $D^0 \to K^- \pi^+ \pi^- \pi^+$ \\ \hline
  $D^0 \to K^+ \pi^- \pi^0$       & $D^0 \to K^- \pi^+ \pi^0$       \\ \hline
  $D^0 \to K^+ \pi^- \pi^0 \pi^0$ & $D^0 \to K^- \pi^+ \pi^0 \pi^0$ \\ \hline
  $D^0 \to K^+ \pi^- \eta$        & $D^0 \to K^- \pi^+ \eta$        \\ \hline
  $D^0 \to K^+ \pi^- \pi^0 \eta$  & $D^0 \to K^- \pi^+ \pi^0 \eta$  \\ \hline
  \hline
\end{tabular}
\label{BKG1}
  \end{table}

\begin{figure}[htbp]
  \centering
  \includegraphics[width=0.70\textwidth]{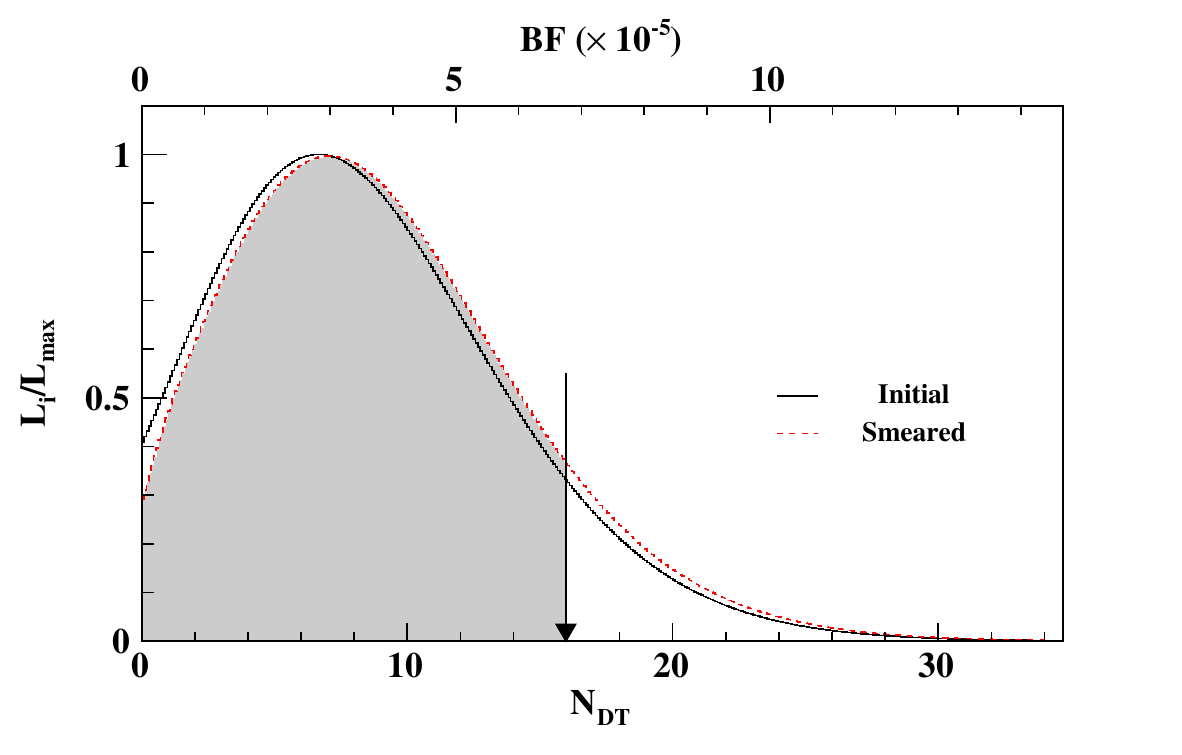}
  \caption{Distribution of the normalized likelihood versus the signal yield $N_{\rm DT}$ and the normalized BF of $D^0\to K^+\pi^-\pi^0\eta$.
   The dashed red and solid black curves show the results obtained with and without incorporating the systematic uncertainty, respectively. The black arrow shows the value corresponding to the 90\% confidence level.}
  \label{fig:upper}
\end{figure}

No significant signal of $D^0\to K^+\pi^-\pi^0\eta$ is observed.
The BF upper limit is set to be $6.8\times 10^{-5}$ at the 90\% confidence level, using the Bayesian approach~\cite{Feldman:1997qc,Stenson:2006gwf,Convery:2003af,BESIII:2021drk} after incorporating the systematic uncertainty
via~\cite{Stenson:2006gwf}
\begin{equation}
  L(\mathcal{B}) \propto \int_0^1 L\left(\mathcal{B} \frac{\epsilon}{\epsilon_0}\right) \exp \left[\frac{-\left(\epsilon / \epsilon_0-1\right)^2}{2\left(\sigma_\epsilon\right)^2}\right] d \epsilon ,
\end{equation}
where $L(\mathcal{B})$ is the likelihood distribution as a function of $\mathrm{BF}$, $\epsilon$ is the efficiency including systematic effects, while $\epsilon_0$ is the nominal MC-estimated efficiency.
The upper limits on the product of the BFs at the 90\% confidence level are obtained by integrating $L(\mathcal{B})$ from zero to 90\% of the total area of the curve.
The distribution of the normalized likelihood versus the assumed BF is shown in Fig.~\ref{fig:upper}.

\section{Analysis of the DCS $\mathbf{D^+}$ decays}

The single-tag~(ST) $D^-$ candidates are selected by reconstructing a $D^-$ in the hadronic decay modes
$D^- \to K^{+}\pi^{-}\pi^{-}$,
$D^- \to K^0_{S}\pi^{-}$ and $D^-\to K^{+}\pi^{-}\pi^{-}\pi^{0}$.
Events in which a signal candidate is reconstructed in the presence of an ST $D^-$ meson
are referred to as DT events.
The BF of the signal decay is determined by
\begin{equation}
  \label{eq:br}
  {\mathcal B}_{\rm sig} = N_{\rm DT}/(N^{\rm tot}_{\rm ST} \, \epsilon_{\rm sig}),
\end{equation}
where
$N^{\rm tot}_{\rm ST}=\sum_i N_{\rm ST}^i$ and $N_{\rm DT}$
are the total yields of the ST and DT candidates in data, respectively.
Here, $N^i_{\rm ST}$ is the ST yield for the tag mode $i$, and
the efficiency $\epsilon_{{\rm sig}}$ for detecting the signal $D^+$ decay
is averaged over the tag mode $i$,
\begin{equation}
  \epsilon_{\text {sig }}=\frac{\sum_i\left(N_{\mathrm{ST}}^i \,  \epsilon_{\mathrm{DT}}^i / \epsilon_{\mathrm{ST}}^i\right)}{N_{\mathrm{ST}}^{\mathrm{tot}}},
\end{equation}
where $N_{\mathrm{ST}}^i$ is the number of ST $D^{-}$mesons for the $i^{\rm th}$ tag mode in data, $\epsilon_{\mathrm{ST}}^i$ is the efficiency of reconstructing the ST mode $i$ (the ST efficiency), and $\epsilon_{\mathrm{DT}}^i$ is the efficiency of finding the tag mode $i$ and the DCS $D^{+}$ decay simultaneously (the DT efficiency).

The selection criteria of $K^\pm$, $\pi^\pm$, $K^0_S$, $\pi^0$ and $\eta$
have been discussed in Sec.~\ref{eventselection}.
Tagged $D^-$ mesons are identified using the energy difference $\Delta E_{\rm tag} \equiv E_{D^-} - E_{\rm beam}$
and the beam-constrained mass $M_{\rm BC}^{\rm tag} \equiv \sqrt{E^{2}_{\rm beam}-|\vec{p}_{D^-}|^{2}}$, where $\vec{p}_{D^-}$ and $E_{D^-}$ are the momentum and the energy of the $D^-$ in the rest frame of the $e^+e^-$ system, respectively.
For each tag mode, if there are multiple candidates in an event,
only the one with the smallest $|\Delta E_{\rm tag}|$ is retained.
The $\Delta E_{\rm tag}$ requirements for the ST $D^-$ candidates are listed in Table~\ref{tab:singletag}; they vary due to differing resolutions.

To extract the yields of ST $D^-$ candidates for each tag mode, binned maximum-likelihood fits are performed on the corresponding $M_{\rm BC}^{\rm tag}$
distributions, following Ref.~\cite{BESIII:2023exq}.
The $D^-$ signal is modeled by the MC-simulated shape convolved with
a double-Gaussian function describing the resolution differences between data and MC simulation.
The combinatorial background shape is described by an ARGUS function~\cite{ARGUS:1990hfq}.
The ST $D^-$ yields in data and the ST efficiencies, estimated by analyzing the inclusive MC sample, are listed in Table~\ref{tab:singletag}.

\begin{figure*}[htbp]\centering
  \includegraphics[width=1.0\linewidth]{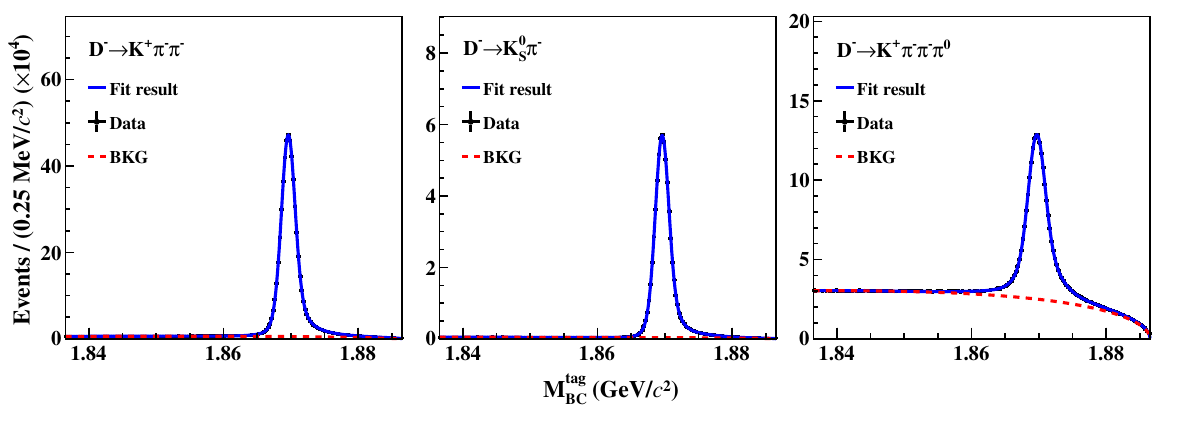}
  \caption{
    Fits to the $M^{\rm tag}_{\rm BC}$ distributions of the ST $D^-$ candidates.
    The points with error bars correspond to the data, the blue curves are the fit results, and the dashed red curves describe the fitted combinatorial background.
  }\label{fig:datafit_Massbc}
\end{figure*}

\begin{table}[htbp]\centering
  \centering
  \caption{The $\Delta E_{\rm tag}$ requirements,
    the yields of the ST $D^{-}$ mesons in data, $N_{\rm ST}$, and the ST efficiencies, $\epsilon_{\rm ST}$. The uncertainties are statistical.
  }\label{tab:singletag}
  \begin{tabular}{c|c|c|c}
    \hline \hline
    Tag mode & $\Delta E_{\rm tag}$ (MeV) & $N_{\rm ST}\,(\times10^3)$  & $\epsilon_{\rm ST}$ (\%)\\
    \hline
    $D^-\to K^{+}\pi^{-}\pi^{-}$ &              $(-25,\,24)$    &$5711.0\pm2.6$    &$52.40\pm0.01$\\
    $D^-\to K^{0}_{S}\pi^{-}$ &                 $(-25,\,26)$    &$667.0\pm0.8$      &$52.60\pm0.01$\\
    $D^-\to K^{+}\pi^{-}\pi^{-}\pi^{0}$ &       $(-57,\,46)$    &$1850.3\pm1.9$    &$25.91\pm0.01$\\
    \hline
    Sum &  &  $8228.1\pm3.3$&  \\
    \hline\hline
    \end{tabular}
\end{table}

\begin{figure*}[htbp]
  \centering
  \includegraphics[width=1.0\textwidth]{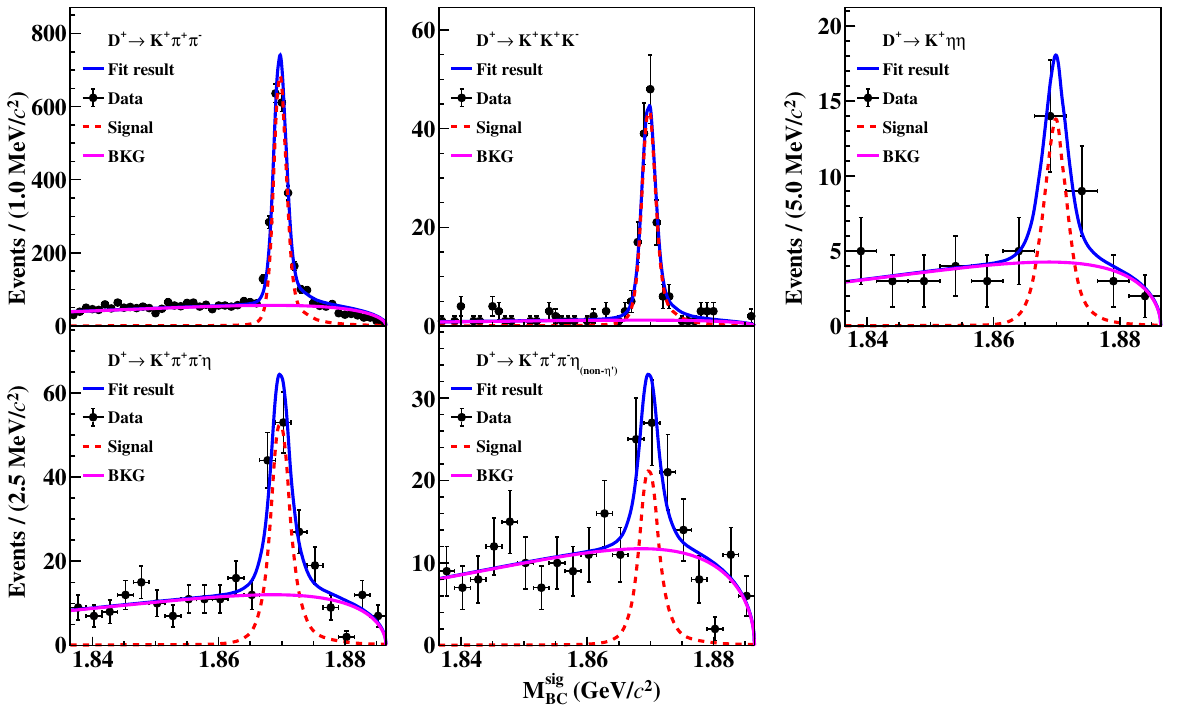}
  \caption{Fits to the $M_{\rm BC}^{\rm sig}$ distributions of the DT candidates in data.
    The points with error bars are data,
    the solid blue curves are the fit results, the dashed red and solid pink curves are the fitted signals and combinatorial backgrounds, respectively.
  }
  \label{dpsigmbcfityidle}
\end{figure*}

The signal $D^+$ decays are selected using the remaining tracks and showers recoiling against the tagged $D^-$ candidates.
In the selection of the $D^+\to K^+\pi^+\pi^-$ and the $D^+\to K^+\pi^+\pi^-\eta$ candidates,
the $\pi^+\pi^-$ combinations are required to be outside the $K_S^0$ mass window, $|M_{\rm\pi^+\pi^-}-0.4977|<0.0300$~GeV/$c^2$, to reject the dominant peaking background from the singly Cabibbo-suppressed decay $D^+\to K^+K_S^0(\to\pi^+\pi^-)$ and $D^+\to K^+K_S^0(\to\pi^+\pi^-)\eta$.
This requirement corresponds to about five standard deviations of the experimental $K_S^0$ mass resolution.
For $D^+\to K^+\pi^+\pi^-\eta$, we also measure the BF of  $D^+\to K^+\left(\pi^{+} \pi^{-} \eta\right)_{{\rm non}-\eta^{\prime}}$ after rejecting the $\eta^\prime$ contribution in the $M_{\pi^+\pi^-\eta}$ invariant mass distribution with the requirement $|M_{\rm\pi^+\pi^-\eta}-0.958|>0.050$~GeV/$c^2$~\cite{BESIII:2020pxp}.

Signal $D^+$ mesons are identified using the energy difference $\Delta E_{\rm sig}$ and the beam-constrained mass $M_{\rm BC}^{\rm sig}$, calculated similarly to the ST side.
For each signal mode, if there are multiple candidates in an event, only the one with the minimum $|\Delta E_{\rm sig}|$ is chosen.
Signal decays are required to satisfy the $\Delta E_{\rm sig}$ requirements listed in Table~\ref{tab:Dp_sigyield}.
The signal yields~($N_{\rm DT}$) are obtained from unbinned maximum likelihood fits to the $M_{\rm BC}^{\rm sig}$ distributions of the DT candidate events in data.
In the fits, the background shapes are described by an ARGUS function and the signal shapes are derived from the signal MC samples.
The results of the fits are shown in Fig.~\ref{dpsigmbcfityidle}.
Table~\ref{tab:Dp_sigyield} gives the fitted DT yields, signal efficiencies, and obtained BFs, and statistical significances.  

\begin{table*}[htpb]
  \centering
  \caption{The $\Delta E_{\rm sig}$ requirements, the DT yield ($N_{\rm DT}$) in data, the signal efficiencies, the obtained BFs, and the statistical significance for each $D^+$ signal decay. The uncertainties are statistical.}
  \renewcommand\arraystretch{1.2}
  \adjustbox{width=0.99\textwidth}{%
    \begin{tabular}{l|c|c|c|c|c}
    \hline\hline
    Signal decay                                               & $\Delta E_{\rm sig}$ (MeV) & $N_{\rm DT}$            & $\epsilon_{\rm sig}~(\%)$ & $\mathcal B_{\rm sig}~(\times10^{-4})$ & Significance ($\sigma$) \\ \hline
$ D^{+} \to  K^{+} \pi^{+} \pi^{-}$&$(-24,23)$&2001.5$\pm$51.6&54.00$\pm$0.20&4.50$\pm$0.12&$>12.0$\\
$ D^{+} \to  K^{+} K^{+} K^{-}$&$(-21,19)$&142.3$\pm$13.1&33.88$\pm$0.17&0.51$\pm$0.05&$>12.0$\\
$ D^{+} \to  K^{+} \pi^{+} \pi^{-} \eta$&$(-28,27)$&96.6$\pm$13.3&19.11$\pm$0.14&1.56$\pm$0.22&9.7\\
$ D^{+} \to  K^{+} \left(\pi^{+} \pi^{-} \eta\right)_{{\rm non}-\eta'}$&$(-28,27)$&38.9$\pm$10.2&17.82$\pm$0.26&0.67$\pm$0.18&4.4\\
$ D^{+} \to  K^{+} \eta \eta$&$(-29,29)$&14.6$\pm$5.6&19.36$\pm$0.13&0.59$\pm$0.23&3.2\\
  \hline \hline
  \end{tabular}%
  }
  \label{tab:Dp_sigyield}
\end{table*}

\section{Systematic uncertainties}

The systematic uncertainties in the BF measurements are described below and listed in Table~\ref{tab:totsys} for each decay mode. The total systematic uncertainty is given by their sum in quadrature.

\subsection{Common systematic uncertainties}

The $e^-$ tracking~(PID) efficiencies
are studied by using as control sample the $e^+e^-\to\gamma e^+ e^-$ events and those for $\mu^-$ by using $e^+e^-\to\gamma \mu^+ \mu^-$ events.
We assign
1.0\% as the systematic uncertainty from the $e^-(\mu^-)$ tracking and PID.
The $K^+$ and $\pi^-$ tracking~(PID) efficiencies are investigated with the DT hadronic $D\bar D$ events $D^0 \to K^-\pi^+$, $K^-\pi^+\pi^0$, $K^-\pi^+\pi^+\pi^-$ versus~$\bar D^0 \to K^+\pi^-$, $K^+\pi^-\pi^0$, $K^+\pi^-\pi^-\pi^+$, and $D^+\to K^- \pi^+\pi^+$ versus~$D^-\to K^+\pi^-\pi^-$.
We assign 0.5\% as the systematic uncertainty of the tracking or PID per $K^\pm$ or $\pi^\pm$.

The $\pi^0$ reconstruction efficiency is studied by using the DT hadronic $D\bar D$  decays 
$\bar D^0\to K^+\pi^-\pi^0$ and $\bar D^0\to K^0_S\pi^0$ tagged by either $D^0\to K^-\pi^+$ or $D^0\to K^-\pi^+\pi^+\pi^-$~\cite{BESIII:2016gbw,BESIII:2016hko}.
Due to the limited $\eta$ sample and a topology similar to the $\pi^0$ sample, the uncertainty in the $\eta$ reconstruction is assigned as equal to that of the $\pi^{0}$ reconstruction: 1.0\% for each $\pi^0$ or $\eta$. 

The systematic uncertainties due to the requirements on $\Delta E$ and $M_{\rm BC}$ for the signal side are studied
by using control samples of the CF decays $D^0\to K^-\pi^+\pi^0$ and $D^0\to K^-\pi^+\pi^+\pi^-$.
The
difference of the efficiencies between data and MC simulation, 0.1\%, is assigned as the systematic uncertainty.

The uncertainties from the limited MC statistics are determined by the signal efficiencies obtained using the signal MC samples.
The uncertainty in MC statistics is assigned by
$\sqrt{(1-\epsilon)/\epsilon}\,/\,\sqrt{N} $,
where $\epsilon$ is the detection efficiency and $N$ is total number of MC events.

The systematic uncertainty associated with the $K^0_S$ veto is assigned by varying the nominal veto mass window by $\pm20$~MeV/$c^2$.
The maximum relative change in the measured BF is not significantly larger than the statistical
uncertainty after considering the correlations between the signal yields. Therefore, this uncertainty is neglected~\cite{Barlow:2002yb}.

For the multi-body DCS $D$ decays, we generate the exclusive MC samples including the possible resonances~($\rho^{0(-)}$, $K^{*0(+)}$, $f_0$ and $\phi$).
The differences of the averaged and nominal signal efficiencies are taken as the systematic uncertainty of the MC model.
An uncertainty of the MC model for $\bar D^0 \to K^+ e^- \nu_e$~\cite{BESIII:2015tql}, 0.1\%, is also considered in the measurement of the $D^0$ DCS decays.

\subsection{Systematic uncertainties for the $\mathbf{D^0}$ decays}

The systematic uncertainties associated with the $U_{\rm miss}$ fit are considered below.
The systematic uncertainty due to the signal shape is determined by varying the width of the smeared Gaussian function by $\pm 1\sigma$.
To estimate the uncertainties from the floating background sources, we examine the re-measured BF by varying the background shape to a Chebychev function in the fit.
The maximum changes of the signal yields are assigned as the systematic uncertainties.
Adding these effects in quadrature gives the systematic uncertainties due to the $U_{\rm miss}$ fit for the DCS $D^0$ decays.

The systematic uncertainty in the $M_{K^+\ell^-}$ requirement is estimated by using the candidates for $\bar D^0\to K^+\ell^-\bar \nu_\ell$ versus the corresponding CF decays.
The differences of the DT efficiencies, 0.1\% and 0.3\%, are assigned as the systematic uncertainties for the requirements of $M_{K^+e^-}$ and $M_{K^+\mu^-}$~\cite{BESIII:2022chs}, respectively.

The systematic uncertainty due to the $E^{\rm max}_{\rm extra\, \gamma}$, $N_{\rm extra}^{\rm charge}$, and $N_{\rm extra}^{\pi^0}$ requirements
is estimated by using a control sample of the corresponding CF decays versus $\bar D^0\to K^+ e^-\bar \nu_e$.
The differences in the acceptance efficiencies between data and MC simulation are taken as the systematic uncertainties.

The uncertainty from the FSR recovery is assigned as 0.3\%, as obtained in Ref.~\cite{BESIII:2015tql}.
The uncertainty in the total number of the $D^0\bar D^0$ pairs
is 1.1\%\cite{BESIII:2024lbn}.
The uncertainties of the quoted BFs  ${\mathcal B}(D^0\to K^-e^+\nu_{e})=(3.549\pm0.026)\%$ and ${\mathcal B}(D^0\to K^-\mu^+\nu_{\mu})=(3.41\pm0.04)\%$ are $0.7\%$ and $1.2\%$.
The sum in quadrature of their contributions after taking into account the weights $\epsilon^\ell_{\rm DT} \times {\mathcal B}(D^0\to K^-\ell^+\nu_\ell)$, $0.6\%$, is taken as the systematic uncertainty for each of the DCS $D^0$ decays.

\subsection{Systematic uncertainties for the $\mathbf{D^+}$ decays}

The uncertainty of the total yield of the ST $D^-$ mesons is assigned as 0.1\% based on varying the signal shape, the background shape and floating the parameters of one smearing Gaussian in the $M^{\rm tag}_{\rm BC}$ fit~\cite{BESIII:2023exq}.

The systematic uncertainties in the $M^{\rm sig}_{\rm BC}$ fit are estimated by changing the signal and background shapes.
The alternative signal shape is chosen from MC candidates where the differences between the true and the reconstructed momentum directions for all tracks agree to within $15^{\circ}$. 
The alternative background shapes are obtained by varying the endpoint of the ARGUS function~\cite{ARGUS:1990hfq}.
The changes of the signal yields relative to the nominal ones are assigned as the systematic uncertainties.

\begin{table*}[htpb]
  \centering
  \caption{Relative systematic uncertainties for the BF measurements, in \%. The ellipsis~($\mathit{...}$) indicate cases where a source is not applicable.}
  \label{tab:totsys}
  % \scalebox{0.9}{%
  \adjustbox{width=\textwidth}{%
    \begin{tabular}{c|cccccc|cccc}
      \hline\hline
      \multicolumn{1}{c|}{\multirow{2}{*}{Source}}                                        & \multicolumn{6}{c|}{$D^0 \to$}  & \multicolumn{4}{c}{$D^+ \to$}                                                                                                                                                                               \\ \cline{2-11}
      \multicolumn{1}{c|}{}                                                               &
      \multicolumn{1}{c|}{$K^{+} \pi^{-}$}                                                &
      \multicolumn{1}{c|}{$K^{+} \pi^{-} \pi^{-}  \pi^{+}$}                               &
      \multicolumn{1}{c|}{$K^{+} \pi^{-} \pi^{0}$}                                        &
      \multicolumn{1}{c|}{$K^{+} \pi^{-} \pi^{0} \pi^{0}$}                                &
      \multicolumn{1}{c|}{$K^{+} \pi^{-} \eta$}                                           &
      $K^{+} \pi^{-} \pi^{0}  \eta$                                                       &
      \multicolumn{1}{c|}{$K^{+} \pi^{+} \pi^{-}$}                                        &
      \multicolumn{1}{c|}{$K^{+} \pi^{+} \pi^{-} \eta$}                                   &
      \multicolumn{1}{c|}{$K^{+} K^{+} K^{-}$}                                            &
      $K^{+}  \eta \eta$                                                                                                                                                                                                                                                                                                                  \\ \hline
      Tracking &
      \multicolumn{1}{c|}{1.8}  &\multicolumn{1}{c|}{2.3} &\multicolumn{1}{c|}{1.8} &\multicolumn{1}{c|}{1.8} & \multicolumn{1}{c|}{1.8} & 1.8 & \multicolumn{1}{c|}{1.1} & \multicolumn{1}{c|}{1.1} & \multicolumn{1}{c|}{1.5} & 0.5 \\ %\hline 1
      PID &
      \multicolumn{1}{c|}{1.8}  &\multicolumn{1}{c|}{2.3} &\multicolumn{1}{c|}{1.8} &\multicolumn{1}{c|}{1.8} & \multicolumn{1}{c|}{1.8} & 1.8 & \multicolumn{1}{c|}{1.1} & \multicolumn{1}{c|}{1.1} & \multicolumn{1}{c|}{1.5} & 0.5 \\ %\hlines 2
      $\eta(\pi^0)$ reconstruction &
      \multicolumn{1}{c|}{...}    &\multicolumn{1}{c|}{...}   &\multicolumn{1}{c|}{1.0} &\multicolumn{1}{c|}{1.4} & \multicolumn{1}{c|}{1.0} & 1.4 & \multicolumn{1}{c|}{...}   & \multicolumn{1}{c|}{1.0} & \multicolumn{1}{c|}{...  } & 1.4 \\ %\hline 3
      $K^0_S$ veto &
      \multicolumn{1}{c|}{...}    &\multicolumn{1}{c|}{Negligible} &\multicolumn{1}{c|}{Negligible} &\multicolumn{1}{c|}{Negligible} & \multicolumn{1}{c|}{...}   & ...   & \multicolumn{1}{c|}{Negligible} & \multicolumn{1}{c|}{Negligible} & \multicolumn{1}{c|}{...  } & ...   \\ %\hline 4
      $U_{\rm miss}$ fit &
      \multicolumn{1}{c|}{1.0}  &\multicolumn{1}{c|}{3.2} &\multicolumn{1}{c|}{1.5} &\multicolumn{1}{c|}{2.4} & \multicolumn{1}{c|}{1.1} & 0.7 & \multicolumn{1}{c|}{...}   & \multicolumn{1}{c|}{...  } & \multicolumn{1}{c|}{...  } & ...   \\ %\hline 5
      $M^{\rm sig}_{\rm BC}$ fit &
      \multicolumn{1}{c|}{...}    &\multicolumn{1}{c|}{...}   &\multicolumn{1}{c|}{...}   &\multicolumn{1}{c|}{...}   & \multicolumn{1}{c|}{...}   & ...   & \multicolumn{1}{c|}{0.5} & \multicolumn{1}{c|}{0.5} & \multicolumn{1}{c|}{0.5} & 0.5 \\ %\hline 6
      $\Delta E_{\rm sig}$ requirement &
      \multicolumn{1}{c|}{0.1}  &\multicolumn{1}{c|}{0.1} &\multicolumn{1}{c|}{0.1} &\multicolumn{1}{c|}{0.1} & \multicolumn{1}{c|}{0.1} & 0.1 & \multicolumn{1}{c|}{0.1} & \multicolumn{1}{c|}{0.1} & \multicolumn{1}{c|}{0.1} & 0.1 \\ %\hline 7
      $M^{\rm sig}_{\rm BC}$ requirement &
      \multicolumn{1}{c|}{0.1}  &\multicolumn{1}{c|}{0.1} &\multicolumn{1}{c|}{0.1} &\multicolumn{1}{c|}{0.1} & \multicolumn{1}{c|}{0.1} & 0.1 & \multicolumn{1}{c|}{...}   & \multicolumn{1}{c|}{...  } & \multicolumn{1}{c|}{...  } & ...   \\ %\hline 8
      $M_{K^+\ell^-}$ requirement &
      \multicolumn{1}{c|}{0.3}  &\multicolumn{1}{c|}{0.3} &\multicolumn{1}{c|}{0.3} &\multicolumn{1}{c|}{0.3} & \multicolumn{1}{c|}{0.3} & 0.3 & \multicolumn{1}{c|}{...}   & \multicolumn{1}{c|}{...  } & \multicolumn{1}{c|}{...  } & ...   \\ %\hline 9
      $E^{\rm max}_{\rm extra\,\gamma}\&N^{\pi^0}_{\rm extra}\&N^{\rm track}_{\rm extra}$ &
      \multicolumn{1}{c|}{0.4}  &\multicolumn{1}{c|}{0.6} &\multicolumn{1}{c|}{0.1} &\multicolumn{1}{c|}{0.6} & \multicolumn{1}{c|}{0.1} & 0.6 & \multicolumn{1}{c|}{...}   & \multicolumn{1}{c|}{...  } & \multicolumn{1}{c|}{...  } & ...   \\ %\hline 10
      MC model &
      \multicolumn{1}{c|}{0.1}  &\multicolumn{1}{c|}{1.8} &\multicolumn{1}{c|}{0.1} &\multicolumn{1}{c|}{4.8} & \multicolumn{1}{c|}{6.4} & 5.6 & \multicolumn{1}{c|}{7.6} & \multicolumn{1}{c|}{1.9} & \multicolumn{1}{c|}{0.3} & 1.6 \\ %\hline 11
      MC statistics &
      \multicolumn{1}{c|}{0.3}  &\multicolumn{1}{c|}{0.5} &\multicolumn{1}{c|}{0.4} &\multicolumn{1}{c|}{0.7} & \multicolumn{1}{c|}{0.4} & 0.8 & \multicolumn{1}{c|}{0.2} & \multicolumn{1}{c|}{0.4} & \multicolumn{1}{c|}{0.3} & 0.4 \\ %\hline 12
      FSR recovery &
      \multicolumn{1}{c|}{0.3}  &\multicolumn{1}{c|}{0.3} &\multicolumn{1}{c|}{0.3} &\multicolumn{1}{c|}{0.3} & \multicolumn{1}{c|}{0.3} & 0.3 & \multicolumn{1}{c|}{...}   & \multicolumn{1}{c|}{...  } & \multicolumn{1}{c|}{...  } & ...   \\ %\hline
      $N_{D^0\bar D^0}$ &
      \multicolumn{1}{c|}{1.1}  &\multicolumn{1}{c|}{1.1} &\multicolumn{1}{c|}{1.1} &\multicolumn{1}{c|}{1.1} & \multicolumn{1}{c|}{1.1} & 1.1 & \multicolumn{1}{c|}{...}   & \multicolumn{1}{c|}{...  } & \multicolumn{1}{c|}{...  } & ...   \\ %\hline
      $N_{\rm ST}$ &
      \multicolumn{1}{c|}{...}    &\multicolumn{1}{c|}{...}   &\multicolumn{1}{c|}{...}   &\multicolumn{1}{c|}{...}   & \multicolumn{1}{c|}{...}   & ...   & \multicolumn{1}{c|}{0.1} & \multicolumn{1}{c|}{0.1} & \multicolumn{1}{c|}{0.1} & 0.1 \\ %\hline
      Quoted $\mathcal  B_{D^0 \to K^- \ell^+ \nu_e}$ &
      \multicolumn{1}{c|}{0.6}  &\multicolumn{1}{c|}{0.6} &\multicolumn{1}{c|}{0.6} &\multicolumn{1}{c|}{0.6} & \multicolumn{1}{c|}{0.6} & 0.6 & \multicolumn{1}{c|}{...}   & \multicolumn{1}{c|}{...  } & \multicolumn{1}{c|}{...  } & ...   \\ %\hline
      Quoted  $\mathcal B_{\eta(\pi^0) \to \gamma \gamma}$ &
      \multicolumn{1}{c|}{...}    &\multicolumn{1}{c|}{...}   &\multicolumn{1}{c|}{...}   &\multicolumn{1}{c|}{...}   & \multicolumn{1}{c|}{0.5} & 0.5 & \multicolumn{1}{c|}{...}   & \multicolumn{1}{c|}{0.5} & \multicolumn{1}{c|}{...  } & 1.0 \\ \hline
      Total   &
      \multicolumn{1}{c|}{3.1}  &\multicolumn{1}{c|}{5.1} &\multicolumn{1}{c|}{3.4} &\multicolumn{1}{c|}{6.3} & \multicolumn{1}{c|}{7.2} & 6.6 & \multicolumn{1}{c|}{7.8} & \multicolumn{1}{c|}{2.8} & \multicolumn{1}{c|}{2.2} & 2.5 \\ \hline   
    \hline
    \end{tabular}%
  }
\end{table*}

\section{Summary}

By analyzing 20.3~fb$^{-1}$ of $e^+e^-$ collision data collected at $\sqrt{s} = 3.773$~GeV with the BESIII detector, we obtain measurements of several DCS $D^0$ and $D^+$ decays.
Table~\ref{totresult} summarizes the BFs measured in this work, the PDG values of the corresponding DCS and CF decays, the individual DCS/CF ratios and the ratios in unit of $\tan^{4}{\theta}_{C}$.
The decays
$D^0 \to K^+\pi^-\eta$,
$D^0 \to K^+\pi^-\pi^0\eta$,
$D^+\to K^+\eta\eta$,
$D^{+} \to  K^{+} \pi^{+} \pi^{-} \eta$,
and
$D^{+} \to  K^{+} \left(\pi^{+} \pi^{-} \eta\right)_{{\rm non}-\eta^{\prime}}$
are investigated for the first time.
For the known decays 
$D^0 \to K^+ \pi^-$,
$D^0 \to K^+ \pi^- \pi^- \pi^+$,
$D^+ \to K^+ \pi^+ \pi^-$,
and $D^+ \to K^+ K^+ K^-$, the BFs measured in this work are consistent with world average values.
In the future, amplitude analyses of the multi-body DCS $D$ decays with larger data samples taken by BESIII~\cite{Ke:2023qzc,BESIII:2020nme,Li:2021iwf} and the Super Tau-Charm Facility~\cite{Achasov:2023gey} will be able to extract the BFs of the intermediate two-body $D$ decays.  This will help to further explore quark SU(3)-flavor symmetry and its breaking effects, and potentially improve theoretical predictions of $\mathit{CP}$ violation in hadronic $D$ decays~\cite{Saur:2020rgd}.

\begin{table*}[htbp]
  \centering
  \caption{The DCS BFs measured in this work, the PDG values of the corresponding DCS and CF decays, and the new DCS/CF BF ratios, given both directly and in units of $\tan^{4}{\theta}_{C}$.
}
  \renewcommand\arraystretch{1.3}
  \adjustbox{width=\textwidth}{%
    \begin{tabular}{l|c|c|c|c|c}
      \hline \hline
      Signal decay                                               & $\mathcal B_{\rm DCS}^{\rm This~work}~(\times10^{-4})$ & $\mathcal B_{\rm DCS}^{\rm PDG}~(\times10^{-4})$ & $\mathcal B_{\rm CF}^{\rm PDG}~(\times10^{-2})$ & $\mathcal B_{\rm DCS}^{\rm This~work}/\mathcal B_{\rm CF}~(\%) $ & $\times \tan^{4}{\theta}_{C}$ \\ \hline
$D^0 \to  K^{+} \pi^{-}$&$1.30 \pm 0.09 \pm 0.04$&$1.50 \pm 0.07$&$3.947 \pm 0.030$&$0.328 \pm 0.027$&$1.14 \pm 0.09$\\ %\hline
$D^0 \to  K^{+} \pi^{-} \pi^{-}  \pi^{+}$&$2.38 \pm 0.19 \pm 0.12$&$2.65 \pm 0.06$&$8.22 \pm 0.14$&$0.289 \pm 0.028$&$1.00 \pm 0.10$\\ %\hline
$D^0 \to  K^{+} \pi^{-} \pi^{0}$&$3.06 \pm 0.21 \pm 0.10$&$3.06 \pm 0.16$&$14.4 \pm 0.6$&$0.212 \pm 0.021$&$0.74 \pm 0.07$\\ %\hline
$D^0 \to  K^{+} \pi^{-} \pi^{0}  \pi^{0}$&$1.40 \pm 0.27 \pm 0.09$&$<3.6$&$8.86 \pm 0.23$&$0.158 \pm 0.036$&$0.55 \pm 0.12$\\ %\hline
$D^0 \to  K^{+} \pi^{-} \eta$&$1.04 \pm 0.16 \pm 0.08$&$-$&$1.88 \pm 0.05$&$0.555 \pm 0.092$&$1.93 \pm 0.32$\\ %\hline
$D^0 \to  K^{+} \pi^{-} \pi^{0}  \eta$&$<0.7$&$-$&$0.449 \pm 0.027$&$< 1.78 $&$< 6.19 $\\ %\hline
\hline
$ D^{+} \to  K^{+} \pi^{+} \pi^{-}$&$4.50 \pm 0.12 \pm 0.35$&$4.91 \pm 0.09$&$9.38 \pm 0.16$&$0.480 \pm 0.019$&$1.67 \pm 0.07$\\ %\hline
$ D^{+} \to  K^{+} \pi^{+} \pi^{-} \eta$&$1.56 \pm 0.22 \pm 0.04$&$-$&$-$&$-$&$-$\\ %\hline
$ D^{+} \to  K^{+} \left(\pi^{+} \pi^{-} \eta\right)_{{\rm non}-\eta'}$&$0.67 \pm 0.18 \pm 0.02$&$-$&$0.135 \pm 0.012$&$5.0 \pm 1.4$&$17.3 \pm 4.8$\\ %\hline
$ D^{+} \to  K^{+} K^{+} K^{-}$&$0.51 \pm 0.05 \pm 0.01$&$0.614 \pm 0.011$&$-$&$-$&$-$\\ %\hline
$ D^{+} \to  K^{+} \eta \eta$&$0.59 \pm 0.23 \pm 0.02$&$-$&$-$&$-$&$-$\\ %\hline
            \hline
      \hline
    \end{tabular}%
  }
  \label{totresult}
\end{table*}

\section{Acknowledgement}
The BESIII Collaboration thanks the staff of BEPCII and the IHEP computing center for their strong support. This work is supported in part by National Key R\&D Program of China under Contracts Nos. 2023YFA1606000; National Natural Science Foundation of China (NSFC) under Contracts Nos. 12035009, 11875170, 12105076, 11635010, 11735014, 11935015, 11935016, 11935018, 12025502, 12035013, 12061131003, 12192260, 12192261, 12192262, 12192263, 12192264, 12192265, 12221005, 12225509, 12235017, 12361141819; the Chinese Academy of Sciences (CAS) Large-Scale Scientific Facility Program; the CAS Center for Excellence in Particle Physics (CCEPP); Joint Large-Scale Scientific Facility Funds of the NSFC and CAS under Contract No. U1832207; 100 Talents Program of CAS; The Institute of Nuclear and Particle Physics (INPAC) and Shanghai Key Laboratory for Particle Physics and Cosmology; German Research Foundation DFG under Contracts Nos. FOR5327, GRK 2149; Istituto Nazionale di Fisica Nucleare, Italy; Knut and Alice Wallenberg Foundation under Contracts Nos. 2021.0174, 2021.0299; Ministry of Development of Turkey under Contract No. DPT2006K-120470; National Research Foundation of Korea under Contract No. NRF-2022R1A2C1092335; National Science and Technology fund of Mongolia; National Science Research and Innovation Fund (NSRF) via the Program Management Unit for Human Resources \& Institutional Development, Research and Innovation of Thailand under Contracts Nos. B16F640076, B50G670107; Polish National Science Centre under Contract No. 2019/35/O/ST2/02907; Swedish Research Council under Contract No. 2019.04595; The Swedish Foundation for International Cooperation in Research and Higher Education under Contract No. CH2018-7756; U. S. Department of Energy under Contract No. DE-FG02-05ER41374.

\bibliographystyle{JHEP}
\bibliography{refer}

%\clearpage
%\input{BESIIIauthors}

\end{document}